\documentclass[prl,amsmath,aps,twocolumn,superscriptaddress,letterpaper]{revtex4}
\usepackage{xspace,units,subfigure}
\usepackage{float}
\usepackage{amsmath}
\usepackage{amssymb}
\usepackage[normalem]{ulem}
\usepackage[pdftex]{graphicx}

\begin{document}


\title{Memory Effects in the Electron Glass} 
\author{Yasmine Meroz}
\email{yasmine.meroz@weizmann.ac.il}
\author{Yuval Oreg}
\author{Yoseph Imry}
\affiliation{Department of Condensed Matter Physics, Weizmann Institute of Science, Rehovot, 76100, Israel}

\begin{abstract}
We investigate theoretically the slow non-exponential relaxation
dynamics of the electron glass out of equilibrium, where a sudden change
in carrier density reveals interesting memory effects. The
self-consistent model of the dynamics of the occupation numbers in the system
successfully recovers the general behavior found in experiments. Our
numerical analysis is consistent with both the expected logarithmic
relaxation and our understanding of how increasing disorder or
interaction slows down the relaxation process, thus yielding a
consistent picture of the electron glass. We also present a
novel finite size {\it ``domino''} effect where the connection to the leads
affects the relaxation process of the electron glass in mesoscopic
systems.
This effect speeds up the relaxation process, and even reverses the
expected effect of interaction; stronger interaction then leading to a
faster relaxation.
\end{abstract}

\maketitle

\newcommand{\be}{\begin{equation}}
\newcommand{\ee}{\end{equation}}

Characteristic signatures of glassy systems are slow non-exponential
relaxations and nonergodic dynamics which result in aging and memory
effects. Glassy dynamics are ubiquitous, appearing in varied systems
ranging from colloids and bacteria to spins and vortices in
superconductors. The electron glass behavior has been experimentally observed
through nonergodic transport properties of Anderson insulators at low
temperatures for different systems such as amorphous
semiconductors~\cite{Ben-Chorin1991, Ben-Chorin1993, Ovadyahu2008,
  Ovadyahu1997, Vaknin1998, Ovadyahu2007} and granular
metals~\cite{Grenet2007, Grenet2010, Grenet2012}. Exciting the
electron glass from equilibrium by a sudden change of the Fermi level,
i.e., by changing the gate voltage ($V_g$) causes the conductance $G$ of the system to
increase, irrespective of whether the Fermi level was raised or
lowered~\cite{Ben-Chorin1991, Ben-Chorin1993}; (a phenomenon termed the
{\it anomalous field effect}). The excess conductance relaxes back to
equilibrium logarithmically~\cite{Orlyanchik2004}, typically taking
anywhere between minutes and days. This
slow relaxation was attributed to the (unscreened) Coulomb
interaction~\cite{Vaknin1998} which has been the basis of a (local) mean-field
treatment in some theoretical models~\cite{Thouless1977,
  Grunewald1982, Amir2011}. 

For an anomalous field effect, measuring the dependence of the
conductance $G$ on $V_g$ at a fast enough scan rate will reveal a
symmetric component, or dip, in addition to the underlying linear
trend. The dip appears around the Fermi level (i.e $V_g$ at
which the system has been let to equilibrate), and is related to the
{\it Coulomb gap}. This is a soft gap in the density of states (DOS) that results
from the unscreened Coulomb interactions, and close to the Fermi level
takes the form~\cite{Pollak1970, Efros1975, Efros1976}:
\be\label{eq:width_CG}
g(E) \propto \frac{\kappa^d}{e^{2d}}|E|^{d-1},
\ee
where $d$ is the dimension and $\kappa$ the permittivity of the system.

\begin{figure}[t]
\begin{centering}
\includegraphics[width=0.5\textwidth,]{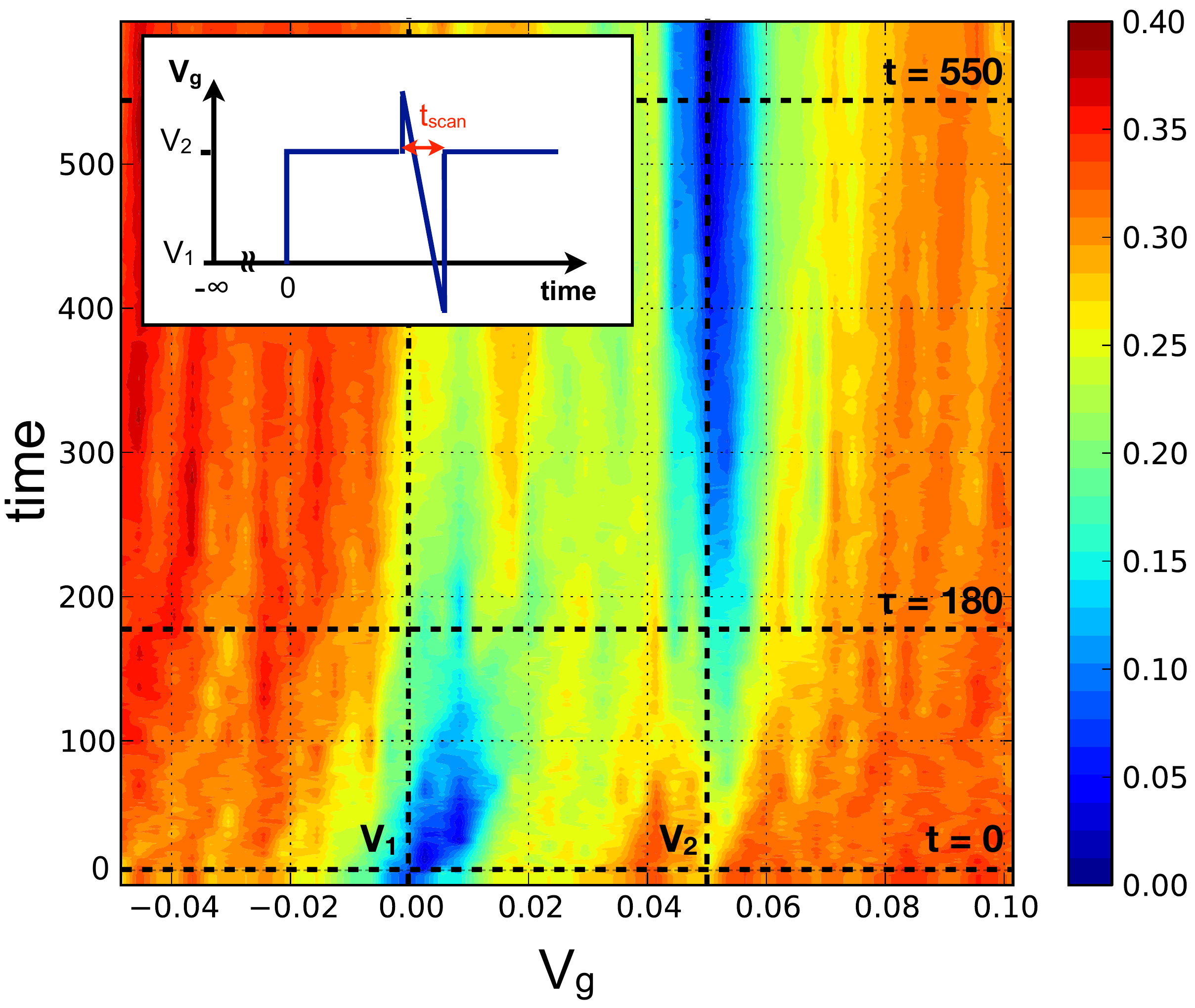}
\par\end{centering}
\caption{(color online) 
Normalized time evolution of the DOS as a function of $V_g$. Cooler colors
  represent a lower DOS. Following the protocol shown in the inset: before $t=0$, the Coulomb gap
  is centred around $V_1 = 0$, at $t=0$ $V_g$ is switched
  to $V_2 = 0.05$ and the original gap slowly disappears, while the new
  gap around $V_2=0.05$ slowly forms. At time $\tau\sim180$ the two dips
  have approximately the same depth. Conductance scans in
  Fig.~\ref{fig:GV} represent times marked here with dashed
  lines. The slight swerve of the fading dip may be due to an effect of
  the leads.
\label{fig:DOS_t}}
\end{figure}

The two-dip
experiment (TDE) is a useful experimental protocol~\cite{Vaknin1998,
  Ovadyahu1997} which probes the memory, dynamics and timescales of the
system, and is described in the inset of Fig.~\ref{fig:DOS_t}: the
system equilibrates at a given $V_g = V_1$, and is then excited by
switching to a new value $V_g = V_2$. Fast scanning measurements of 
$G(V_g)$ expose the equilibration
process, namely the original dip created around $V_1$ gradually disappears while a new dip forms around $V_2$. A time $\tau$ is defined as the time at which
the two dips are of the same depth, and may be associated with a
characteristic relaxation time of the system~\cite{Vaknin1998}, see
also~\cite{Grenet2012}.

In this manuscript we report for the first time a complete theoretical
framework for describing the relaxation dynamics of an electron glass
far from equilibrium. Not only do we numerically recover the general
behavior of the TDE, we also reproduce the expected dependence of
$\tau$ and the width of the dip $\Gamma$ on physical parameters such as
the localization length of the electrons $\xi$ and the permittivity
$\kappa$ (which sets the scale of the interaction strength).
The model we use is based on full equations of 
a local mean-field approach, that to date has only been
used in a linearized form close to equilibrium. We also
report for the first time an important finite size ''domino'' effect, due to the leads, which may greatly
affect the relaxation process in mesoscopic systems and ones with
strong interactions.

We give here the outline of the model originally based
on the picture of a compensated semiconductor, the details
can be found elsewhere~\cite{Monroe1987, Amir2008}. We consider $N$ localized states
with structural and energetic disorder, i.e., each site $i$ has a random position $r_i$ and
a random on-site energy $\epsilon_i$ from the range $[-W/2, W/2]$. The
system consists of $M<N$ Anderson-localized electrons whose transport
is due to phonon assisted hopping from one site to another. The electrons interact via an
unscreened Coulomb potential. We use a local mean-field approach where the
potential energy at site $i$ is given by:
\be\label{eq:E2}
E_i = \epsilon_i + \frac{e^2}{\kappa}\sum_{j\neq i}^{N}{\frac{n_j}{r_{ij}}}.
\ee
Here 
$r_{ij} = |r_i-r_j|$
and $n_i$ is the mean occupation number on site $i$, i.e., $0 \leq
n_i\leq 1$. The transition rate of this tunneling event may be calculated
in the case of weak electron-phonon coupling, where it may be treated
as a perturbation, and takes the form:
\be\label{eq:Gij}
\Gamma_{ij} \propto n_i(1-n_j)e^{-r_{ij}/\xi}[1+\mathcal{N}(|E_{ij}|)]
\ee
where $\xi$ is the localization
length of the electron states, $\mathcal{N}$ is the Bose-Einstein
ditsribution, and $E_{ij} = E_j  - E_i$. For transitions to a higher energy ($E_j > E_i$), the
square brackets are replaced with just $\mathcal{N}(|E_{ij}|)$.
Electrons do not need phonons to hop elastically from a site to one of the leads,
thus the transition rates to the right and left leads take the form 
$\Gamma_{iR} \propto n_i(1-n_R)e^{-(L - x_{i})/\xi}$ and
$\Gamma_{iL} \propto n_i(1-n_L)e^{-x_{i}/\xi}$,
where $n_ R = 1/(1+e^{(E_i - \mu_R)/k_B T} )$ is the Fermi-Dirac
distribution for the difference between the energy at site $i$ and the right
lead held at a chemical potential $\mu_R$ (correspondingly $\mu_L$ for
the left lead). $L$
is the size of the system (distance between the leads), and $x_i$ is the $x$
position of site $i$. 
Given the transition rates we may now write down the coupled,
nonlinear kinetic equations that govern the time evolution of the
occupation numbers at each site:
\be\label{eq:master}
\frac{dn_i}{dt} = \sum_j{\left(\Gamma_{ji} - \Gamma_{ij}\right)}.
\ee
Averaging the solution of Eq.~(\ref{eq:master}) over an ensemble of realizations yields
the evolution of the DOS in time.

We note that previous works~\cite{Amir2008, Amir2009a, Amir2009b} used
a linearized form of the kinetic equations in Eq.~(\ref{eq:master}), expanded
around a local equilibrium state~\cite{Amir2008}. In the case of the
TDE, changing $V_g$ means moving to a different equilibrium. We
are therefore compelled to use the full set of coupled nonlinear equations.

We consider a system of $N=25$ donor sites with half filling, $\xi = 1$, $e^2 = 1$,
$\kappa=25$ and the {\it average} distance between donors is $r_{nn}=1$. The
energy disorder is set to $W=1$, and the leads are kept at $V=0$. The size of the time steps is $dt=0.001$. The dynamics are
calculated for $\sim 2\times 10^4$ different configurations (for each configuration
the uncorrelated positions and on-site energies are randomly chosen).

We numerically solve the kinetic equations in Eq.~(\ref{eq:master})
for this ensemble of realizations, calculating the local mean-field energies
using Eq.~(\ref{eq:E2}) every 15 time steps. At $t=0$ the system starts from equilibrium with $V_1=0$,
exhibiting a Coulomb gap around $V=0$. Following the TDE protocol, at $t=0$ $V_g$
is switched to $V_2=0.05$ (an arbitrarily chosen
number, small compared to the energy disorder). We now calculate
the evolution of the DOS as a function of time. 
We look at the relative change in time of the DOS as done elsewhere~\cite{Vaknin1998,
  Ovadyahu1997}, i.e., $\Delta E / E$ where $\Delta E = E - E_{\text{min}}$, the
difference from the lowest energy in the DOS.

In Fig.~\ref{fig:DOS_t} we show the resulting time evolution of the
DOS as a function of $V_g$ instead of energy, since the experimental mesurements
are done using $V_g$ scans. Note that these scans are assumed to be
conducted fast enough so they do not affect the system. In our case we
literally freeze the time during the scans. This results in a {\it mirror image} of
the DOS as a function of energy, $E \leftrightarrow V_g$, since when one carries out a
measurement at some $V_g$, one effectively raises (lowers) the energy
of the system by $E=V_g$, but probes the system at $E=0$, the
potential at the leads, or the Fermi level $\mu=0$. Therefore effectively one measures the DOS at
$E=-V_g$. 

Fig.~\ref{fig:DOS_t} clearly exhibits the fading out of the original
Coulomb gap around $V_1 = 0$ and the simultaneous formation of the new gap
around $V_2=0.05$. Here the two gaps are approximately equal in
depth at $\tau\sim 180$. A slight swerve of the original dip to higher
values of $V_g$ is apparent as it fades out. This may possibly be due
to an effect of the leads which will be discussed later on.
A slight asymmetry  with respect to $V_g$ at large times may occur as
electrons that get out of the system do not repel any more the
remaining electrons, and hence there is a reduction in the
electrostatic potential. 
We note that if the ratio
between the energy disorder W and the number of sites N is too large,
say $W/N >  0.01$, the level spacing between the sites becomes
significant, and may show up in figures such as Fig.~\ref{fig:DOS_t}
(e.g. the peak around $V_g=0.02$ which starts after $t\approx 300$).

While Fig.~\ref{fig:DOS_t} describes the DOS, experimentally one
measure the conductance between the leads. We calculate the latter
using two methods: 
the first tentative evaluation follows Mott's relation~\cite{Mott1969} between conductance,
temperature and the DOS, through the following
relation:
\be\label{eq:sigma_mott}
G(\mu, T) \propto e^{-\left(\frac{T_0(\mu)}{T}\right)^{\frac{1}{d+1}}} =
e^{-\left(\frac{1}{g(\mu) \xi^d T}\right)^{\frac{1}{d+1}}},
\ee
where $g(\mu)$ is the DOS at the Fermi level. We can emulate the
voltage scans carried out in the TDE
by noting that $V_g$ sets the Fermi level of the system, and by
assuming that within a small enough energy band the DOS can be
considered constant. We can then calculate $\sigma(V_g)$, different
$V_g$ represented by different $\mu$ (Fermi levels) as mentioned above, and
substituting the appropriate value of the calculated DOS in
Eq.~(\ref{eq:sigma_mott})
\footnote{
It is important to note that changing $V_g$ results in a change in
the charge in the system $Q$.The mapping between added charge
and the associated energy change is nonlinear, leading to a cusplike
shape of the dip in the conductance, as opposed to the linear Coulomb gap in
the energy~\cite{Lebanon2005}. This stems from the fact
that the system is between capacitor plates, therefore 
$V_g = Q/C = \frac{1}{C}\int_0^{E}{g(E')}dE' \sim  E^{d}/C$, where
$C$ is the capacitance. Unfortunately our results are too noisy
to tell something about the shape of the resulting dip. We also note
that the DOS is truly zero at the Fermi level for $T=0$, and grows linearly with
$T$~\cite{Levin1987,Mogilyanskii1989}. 
}. 

\begin{figure}[]
\begin{centering}
\includegraphics[width=0.45\textwidth,]{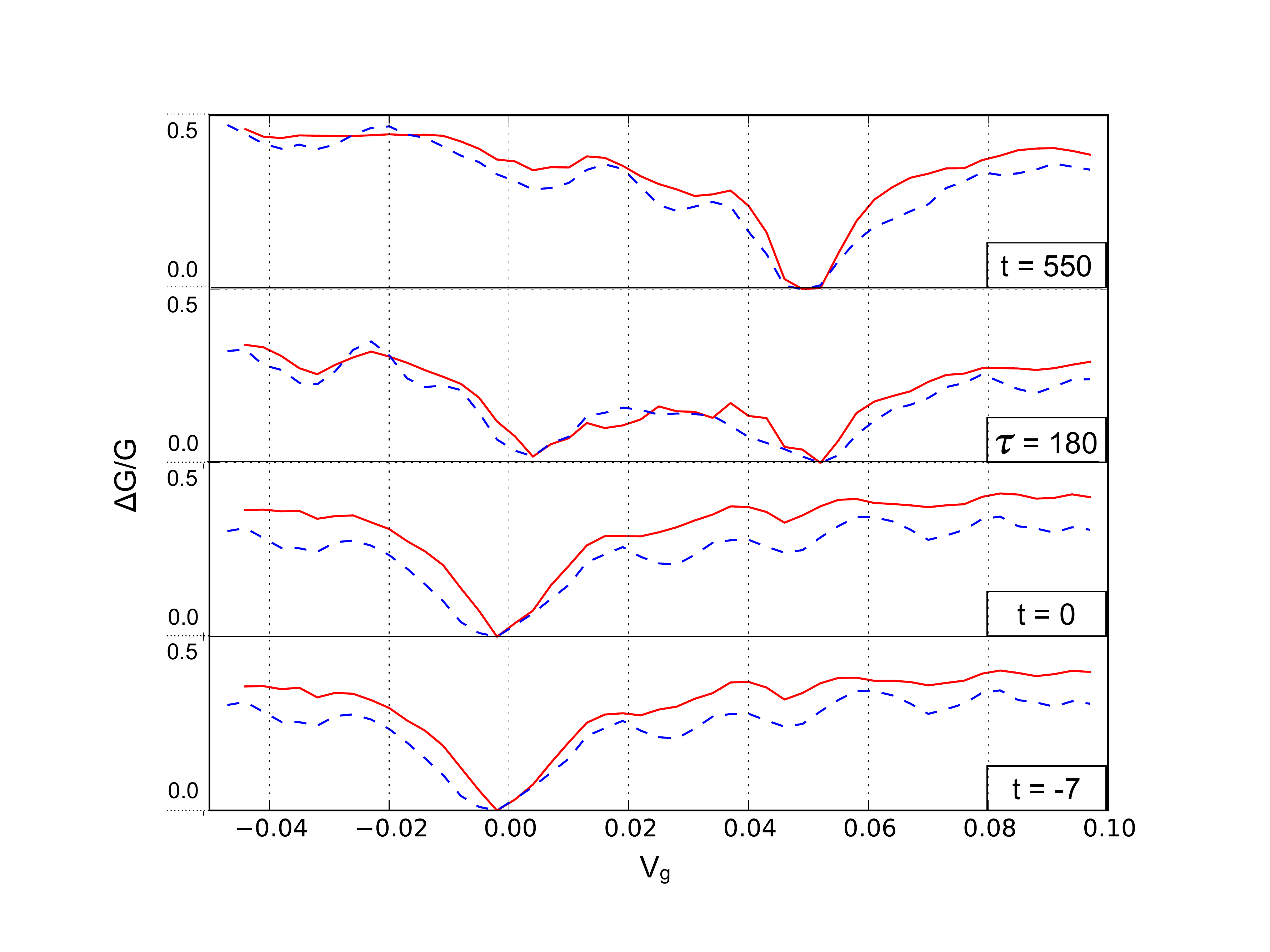}
\par\end{centering}
\caption{(color online) A successful reproduction of the TDE 
  scans $G(V_g)$ at different times, using two
  methods: solid lines use Mott's
  relation, Eq.~(\ref{eq:sigma_mott}), and dashed lines use the Miller-Abrahams random resistor approach,
  Eq.~(\ref{eq:conductance}). The scans are calculated for times marked
  in Fig.~\ref{fig:DOS_t}: (a) $t=-7$, at equilibrium with $V_1 =
  0$, (b) $t=1$, $V_g$ is changed to $V_2=0.05$, (c) $\tau=180$, the time at
  which the two dips are of the same depth, (d) $t=550$, the new dip at $V_2=0.05$
  has formed completely. 
\label{fig:GV}}
\end{figure}

A systematic evaluation of the conductance uses the Miller-Abrahams random
resistor network approach~\cite{Shklovskii1984}, i.e the current between donors $i$ and $j$ is expressed through
the transmission rates: 
$J_{ij} = -e(\Gamma_{ij} - \Gamma_{ji})$. We divide the total current
by the voltage between the leads, chosen to be
$\Delta V = 2 \times 10^{-5}$. The total conductance is the sum of the
conductances between each site and the lead with the lower potential:
\be\label{eq:conductance}
G(V_g) = -\frac{e}{\Delta V}\sum_i{\left(\Gamma_{iR}(V_g) - \Gamma_{Ri}(V_g)\right)}.
\ee
We thus emulate again the contuctance scans by calculating the
conductance in Eq.~(\ref{eq:conductance}) for different $V_g$
(raising or lowering the energy of the system appropriately). 
The conductance scans calculated in both methods, for the same
four times as marked by dashed lines in Fig.~\ref{fig:DOS_t}, are shown
in Fig.~\ref{fig:GV}. 
It turns out that the two methods agree,
successfully reproducing the TDE results. The
data was smoothed by averaging over every three consecutive data
points. Owing to the lack of a prefactor in Eq.~(\ref{eq:sigma_mott}),
the conductances had to be scaled by a multiplication factor to allow a comparison.

\begin{figure}[]
\begin{centering}
\includegraphics[width=0.42\textwidth,]{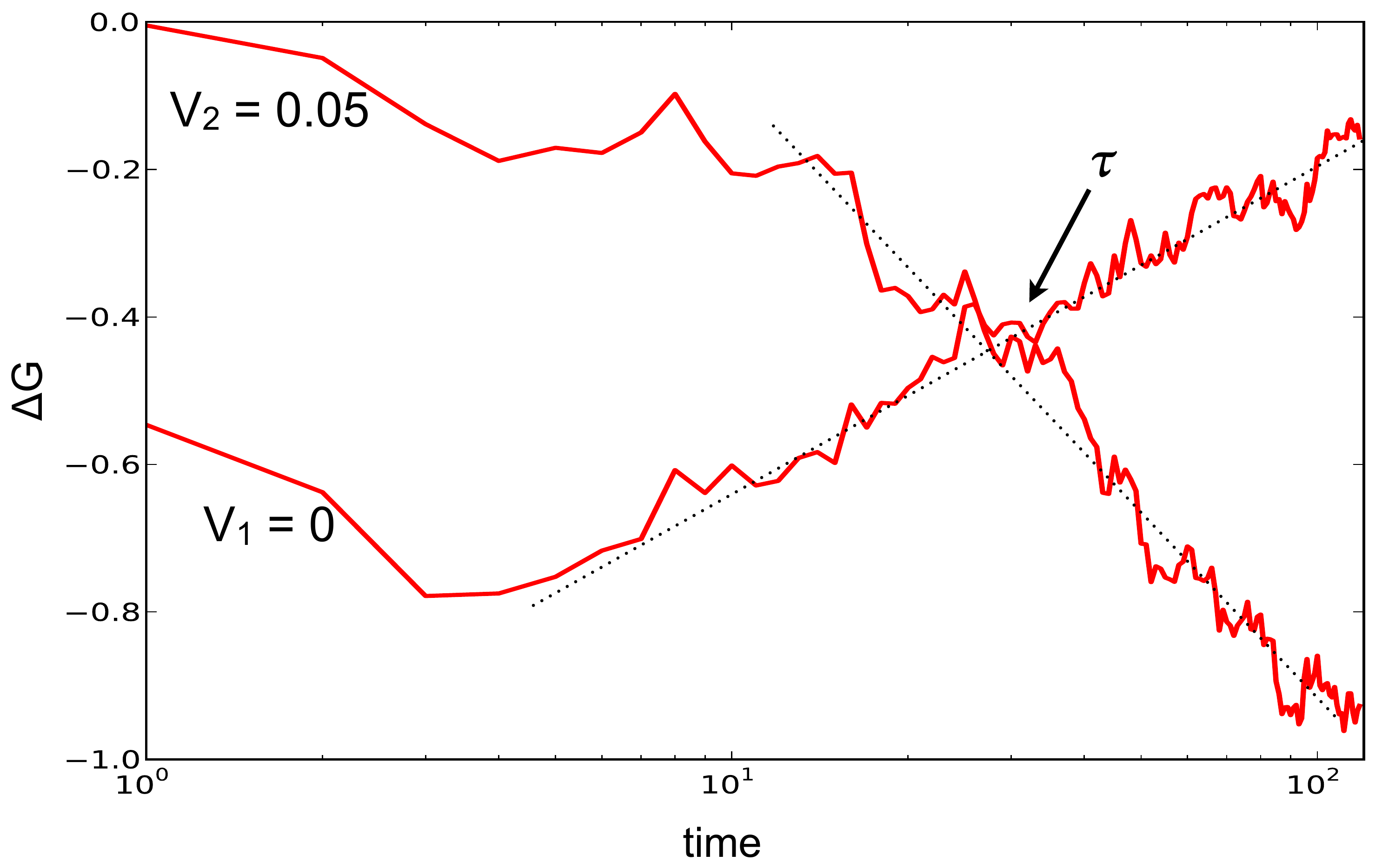}
\par\end{centering}
\caption{The logarithmic relaxation of the two dips, and
  their explicit co-existence. We plot the minima of the conductance dips relative to $G_n$,
  the conductance far from the dip ($\Delta G = G - G_{n}$), as a function of time.
  The curve marked $V_1 = 0$ is the slowly increasing conductance at the original
  dip, and the curve marked $V_2 = 0.05$ represents the slowly
  decreasing conductance at the new forming dip. The time axis is in log
  scale. Dotted lines are guides for the eye. The intersection of the
  lines yields $\tau$, the time at which the two dips have equal depths. 
\label{fig:GV_logt}}
\end{figure}

We further verify our results by ascertaining that the relaxation
process is logarithmic, namely that the conductance at $V_1$ ($V_2$),
i.e., the bottom of the initial dip (new dip) increases (decreases)
logarithmically. This is shown in Fig.~\ref{fig:GV_logt}, where apart
from the expected logarithmic relaxation, one can appreciate the
explicit coexistence of the two dips.

\begin{figure}[t]
\begin{centering}
\includegraphics[width=0.5\textwidth,]{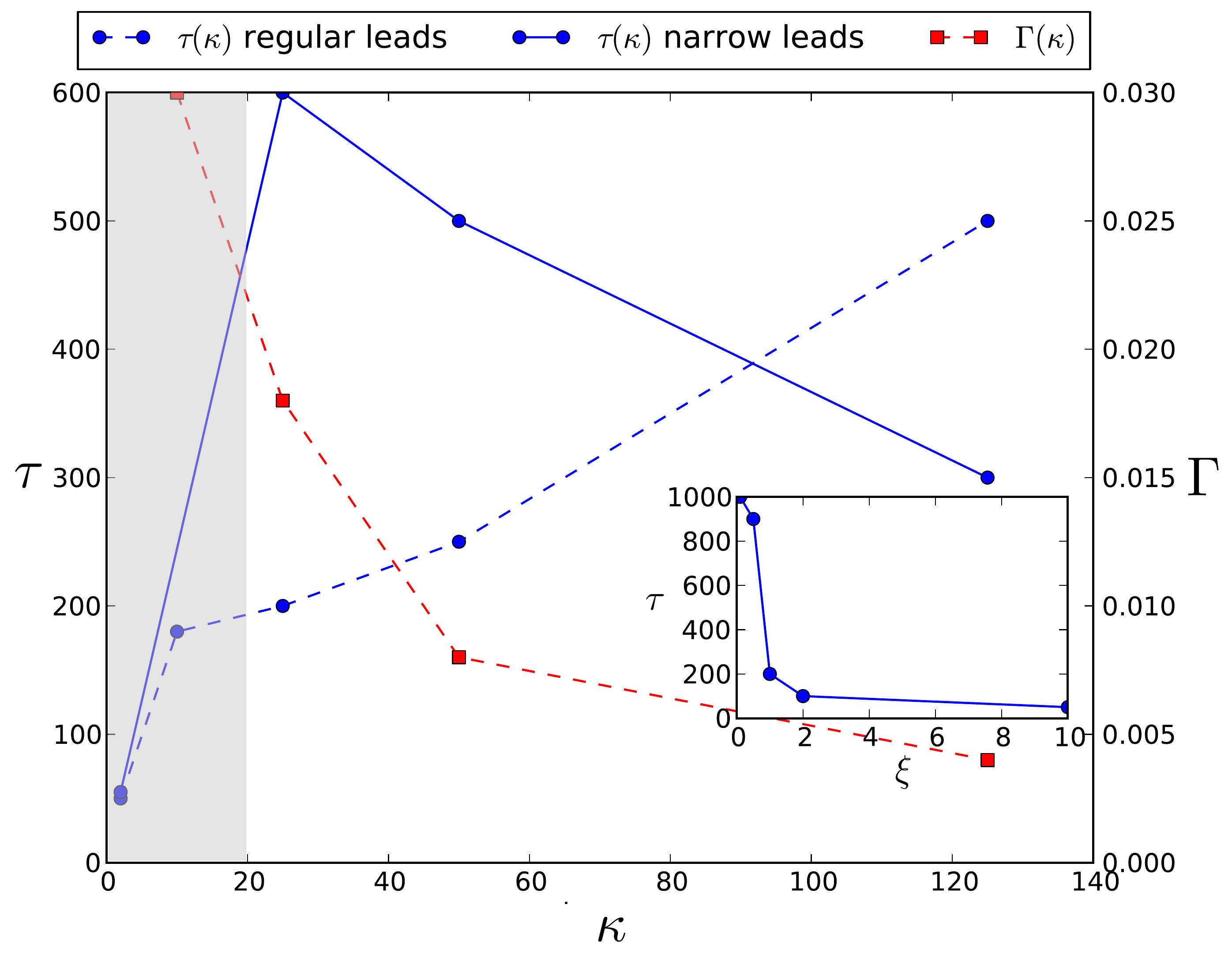}
\par\end{centering}
\caption{(color online) 
  The width of the dip $\Gamma$ taken from the DOS, (squares, scale on right) vs. the
  permittivity $\kappa$, which sets the scale of the interaction
  strength. We find the expected trend of stronger interactions
  leading to a wider dip, as in Eq.~(\ref{eq:width_CG}).
 Qualitative analysis of $\tau$ (circles, scale at left) vs. $\kappa$:
the dashed line represents a square system ($5\times5$ in units of $r_{nn}$)
 and the solid line represents a system with relatively short
  leads (a $2\times12.5$ system). 
Large leads result in a dominant
  {\it domino} effect, where strong
  interactions lead to faster relaxation, as opposed to what is
  expected in a macroscopic glass. With smaller leads the domino effect is less
  dominant, revealing the expected opposite trend of the relaxation
  process. The greyed area marks strong interactions of the same scale
  as the energy disorder (i.e., $e^2/(\kappa r_{nn}) \gtrapprox W$) , where the domino effect is dominant even
  with smaller leads. Inset: $\tau$ as a function of localization
  length $\xi$, showing the expected trend: $\tau$ decreases with $\xi$.
\label{fig:xi_tau}}
\end{figure}

We now proceed to examine the dependence of $\tau$ and the width of
the dip $\Gamma$ on various physical parameters. 
From Eq.~(\ref{eq:width_CG}) we expect $\Gamma$ to be wider for stronger
interactions, as also verified experimentally~\cite{Vaknin1998}. Our model correctly reproduces this
result, shown in Fig.~\ref{fig:xi_tau}.
Secondly, from the transition rates in
Eq.~(\ref{eq:Gij}) we expect $\tau$ to decrease as one increases the
localization length $\xi$. This behavior is indeed recovered by our
model, and is given in the inset of Fig.~\ref{fig:xi_tau}.

\begin{figure}[]
\begin{centering}
\includegraphics[width=0.4\textwidth,]{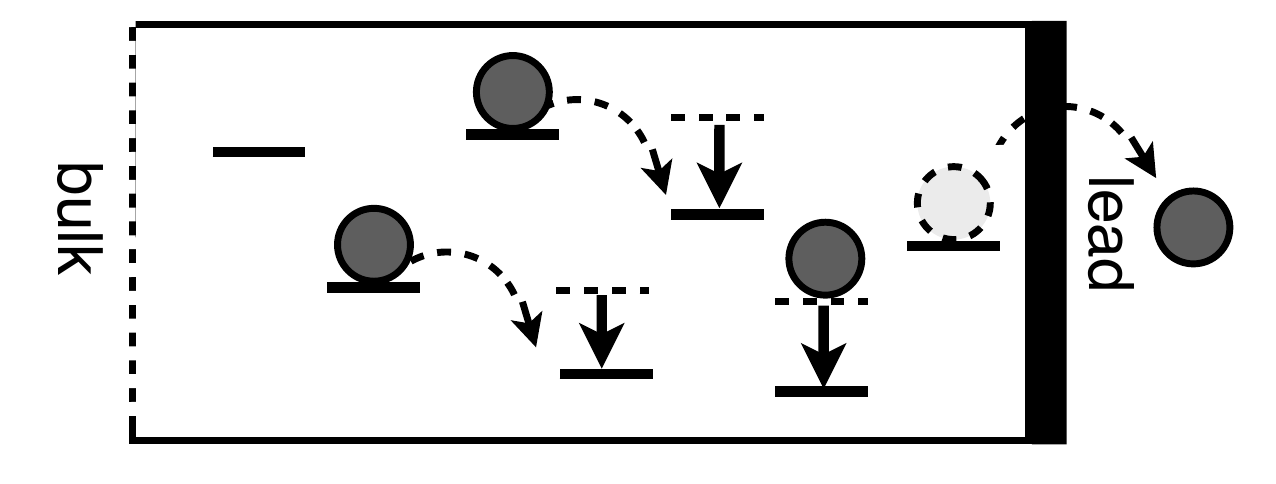}
\par\end{centering}
\caption{A schematic representation of the finite size {\it domino} effect caused by
  the leads on the system. On the
  right is the connection to one of the leads. An electron leaves the system through the
  lead, lowering the energy of other sites by the Coulomb
  energy that is lost, $\frac{e^2}{\kappa r_{ij}}$. Electrons
  farther from the lead will now hop to the sites with newly lowered
  energy. This progressive process of lowering the energy of the sites
  can dramatically speed up the relaxation process, washing out the
  excess electrons through the leads, stronger interactions leading to faster
  relaxations, as opposed to what is expected in glassy systems.
\label{fig:domino}}
\end{figure}

Next let us consider the effect of interaction strength on the
relaxation process: in
glassy systems the relaxation of the system is expected to become {\it
  slower} with stronger interactions, or smaller permittivity $\kappa$. As shown in
Fig.~\ref{fig:xi_tau}, the opposite trend is found. This unexpected
behaviour can be understood when taking into account the fact that
system relaxes through its connection to the leads, and may therefore
be affected by them. Indeed upon raising (lowering) $V_g$,
one raises (lowers) the energy of the sites in the system, raising
(lowering) the initial Fermi energy compared to the potential of the
leads. For simplicity we will discuss the first case of raising $V_g$,
but the argument for lowering $V_g$ is similar.
When the Fermi level is
raised the electrons with excess energy compared to the potential of
the leads leave the system, and the system reorders and relaxes
to its new configuration. We find that when an electron leaves the
system through one of the leads, it causes a {\it domino effect}, 
i.e it leads to a cascade of electron relaxation behind it, thus
speeding up the relaxation in general. 
The effect is explained in Fig.~\ref{fig:domino}.

To verify that this reversal is indeed due to the leads, we minimized
the the effect of the leads by changing the aspect ratio
of the system so as to diminish the relative size of the leads. 
Originally we used a square system $5\times5$ (in units of $r_{nn}$), and
we now used a system $2\times 12.5$, where the leads are of width $2$,
and the area, i.e. the size of the system, is left the same. In this
case we indeed find that relaxation times in general are longer, and
the expected dependence of $\tau$ on the interaction is recovered. For very
strong interactions, of the order of the energy disorder (i.e.,
$e^2/(\kappa r_{nn})\approx W$), the domino
effect becomes dominant again in spite of the relatively small
leads. The results for the square system and the one with narrow leads are compared in 
Fig.~\ref{fig:xi_tau}.
We point out that this effect may also account for
the slight swerve of the original dip to lower energies visible in Fig.~\ref{fig:DOS_t}; in this
approach electrons with higher energy tend to leave the system more quickly
, effectively lowering the original Fermi level. Indeed we noted that
for weaker interactions the swerve is less evident. This issue
deserves further investigation.

We note that our model differs from real systems in a few
aspects: the systems studied experimentally include amorphous and
granular metals, where electrons are thought to tunnel between puddles
of electrons rather than single occupancy sites (as in our model), and
more so it is also expected that simultaneous many-electron
transitions may take place~\cite{Baranovskii1979} 
(disregarded in our model). 
Secondly, in the experiments on amorphous materials one controls the
physical paramters such as $\xi$ and $\kappa$ indirectly through the
carrier density. 
The dependence of $\kappa$ and $\xi$ on $n$ is not universal and may depends
on details of the system (such as distance to screening gates)  that
are not fully  understood.

The electron glass exhibits interesting memory effects
due to ergodicity breaking and aging, which are manifested in the TDE
protocol in the form of complex dynamics of the occupation numbers and
DOS. In this work we successfully reproduced numerically these
experimental results for the first time, by describing the evolution
of the average occupation numbers of sites using kinetic equations, in
a local mean-field approach. Due to the far-from-equilibrium nature of the
problem, we could not use the linear approximation as done before, and
were compelled to solve the full nonlinear coupled equations.
The verification of the logarithmic relaxation of the dip together with our understanding of the
dependence of $\tau$ and $\Gamma$ on the main physical parameters of the system,
$\xi$ and $\kappa$, leaves us with a complete characterization of our model, which
successfully captures the TDE behavior. Moreover we unveiled an
important finite size {\it domino} effect on the relaxation process caused by the leads in the
experimental setup, which not only speeds up the relaxation process in
general, but can also reverse the dependence of the relaxation process
on the interaction strength. This effect should be taken into
consideration when dealing with mesoscopic systems or ones with strong interactions.

{\bf Aknowledgments} We would like to thank Ariel Amir, Thierry
Grenet, Markus M{\"u}ller and Zvi Ovadyahu for fruitful
discussions. This work was supported by the German-Israeli Foundation
(GIF).

\end{document}